\documentclass[aps,prx,reprint,twocolumn,superscriptaddress,nofootinbib,floatfix]{revtex4-2}
 
\usepackage{float}
\usepackage{verbatim}
\usepackage[utf8]{inputenc}
\usepackage[T1]{fontenc}
\usepackage[english]{babel}
\usepackage{dsfont}
\usepackage[normalem]{ulem}
\usepackage{amsmath}
\usepackage{amssymb}
\usepackage{mathrsfs}
\usepackage{amsthm}
\usepackage{mathtools}
\usepackage{wasysym}
\usepackage{breqn}
\usepackage{tabularx}
\usepackage{bbm}
\usepackage{nicefrac}
\usepackage{soul}
\usepackage{stackrel}
\usepackage{braket}
\usepackage{enumitem}
\usepackage{graphicx}
\usepackage{tikz}
\usepackage{mdframed}
\usepackage{booktabs}
\usepackage{siunitx}
\usepackage{fancyhdr}

\usepackage{calc}
\usepackage{accents}

\usepackage{multirow}
\usepackage{array}
\newcolumntype{P}[1]{>{\centering\arraybackslash}p{#1}}

\usepackage[hidelinks]{hyperref}
\usepackage{framed}

\addto\captionsenglish{}

\usepackage{cases}

\addto\captionsenglish{}
\addto\captionsenglish{}

\usepackage{algorithm}



\usepackage{natbib}
\makeatletter
\let\cat@comma@active\@empty
\makeatother

\begin{document}
\title{Improved finite-size analysis for measurement-device-independent
quantum digital signature}

\author{Jia-Li Zhu}
\author{Chun-Hui Zhang}\thanks{chz@njupt.edu.cn}
\author{Yue-Ying Wang}
\author{Qin Wang} \thanks{qinw@njupt.edu.cn}

\affiliation{Institute of quantum information and technology, Nanjing University of Posts and Telecommunications, Nanjing 210003, China}
\affiliation{ ``Broadband Wireless Communication and Sensor Network Technology" Key Lab of Ministry of Education, NUPT, Nanjing 210003, China}
\affiliation{ ``Telecommunication and Networks" National Engineering Research Center, NUPT, Nanjing 210003, China}

\begin{abstract}

\noindent	Quantum digital signatures (QDS), based on the principles of quantum mechanics, provide information-theoretic security, ensuring the integrity, authenticity, and non-repudiation of data transmission. With present QDS protocols, measurement-device-independent QDS (MDI-QDS) can resist all attacks on detections, yet it suffers from finite-size effect. In this work, we present and compare three parameter estimation models for finite-size analysis of two-decoy MDI-QDS. The first model is a commonly used model in previous schemes, and we propose two new models to improve the performance. Subsequently, we perform numerical simulations to evaluate the performance of the three models. The results demonstrate that the proposed methods are less affected by finite-size effect, thereby effectively enhancing the signature rate. This work contributes to the practical development of QDS.
\end{abstract}

\maketitle

\section{Introduction}
With the rapid development of various internet-based services, e.g. e-commerce, digital currencies, the data security is increasing dramatically. Digital signatures, as a means of ensuring the authenticity and integrity of data, are widely used in modern communications \cite{APP1,APP2}. However, security threats in digital signatures do exist all the time. For example, data information is vulnerable to eavesdropping and forging during the signing process. In addition, traditional digital signatures built on the computational complexity of mathematical methods are likely to be cracked by quantum computers in the near future \cite{Eve,Eve1}. This makes digital signatures unreliable in areas with high security requirements, such as military, government, and finance.

Compared to traditional digital signatures, quantum digital signatures (QDS) offer theoretically unconditional security based on the principles of quantum mechanics \cite{QKD1,QKD2,QKD3}, which can guarantee the integrity, authenticity and non-negativity of data transmission. The first QDS protocol was proposed in 2001 \cite{QDS2001}, but it was impractical due to the need for long-term quantum storage and secure quantum channels. Since then, efforts have been made to eliminate these unrealistic technical requirements \cite{SEC1,SEC2,SEC3,SEC4}. Based on various QKD protocols, the performance of QDS has been optimised in terms of security and signature rate. For example, the BB84 protocol-based QDS (BB84-QDS) \cite{BB84QDS} has a high signature rate over short distances, and measurement-device-independent QDS (MDI-QDS) \cite{MDIQDS,MDIQDS2} and twin-field QDS (TF-QDS) \cite{TFQDS,TFQDS2} schemes can significantly extend the transmission distance and defend against all detector-side channel attacks. Additionally, there are many post-processing techniques that can improve the performance of QDS such as random pairing (RP) \cite{PP1}, actively-odd-parity pairing (AOPP) \cite{PP2}, and advantage distillation (AD) \cite{AD1,AD5,AD6}. 

With the development of QDS, the practical deployments of QDS systems have emerged around the world. In 2017, Yin \textit{et al.} conducted an experiment on two-decoy BB84-QDS over a 102 km fiber distance \cite{LD1}. Subsequently, Zhang \textit{et al.} demonstrated a passive decoy-state BB84-QDS system over 40 dB, circumventing the loopholes of active intensity modulation \cite{LD2}. Furthermore, Ding \textit{et al.} reported a one-decoy BB84-QDS system over 280 km, verifying the feasibility of QDS in long-distance communication \cite{LD3}. Beyond long-distance transmission, other research advancements hold substantial significance as well. Regarding high-speed QDS, An \textit{et al.} and Roberts \textit{et al.} respectively realize the gigaherz system of BB84-QDS \cite{HD1} and MDI-QDS \cite{NET3}. In 2023, Yin \textit{et al.} built the first all-in-one quantum secure network integrating information-theoretically secure communication, digital signature, secret sharing and conference key agreement \cite{NET2}. In 2024, Du \textit{et al.} verified a chip-based QDS network, achieving a maximum signature rate of 0.0414 times per second for a 1 Mbit file over fiber distances of up to 200 km \cite{Chip}. In addition, QDS has also progressed to multi-user \cite{MU1,MU2,MU3}, single-photon source \cite{SPS}, free space \cite{FT1}, and metropolitan networks \cite{NET1}.

In practical QDS systems, the signature rate is the number of messages that the system can process and successfully sign in a given time. Since the number of signals that can be exchanged within a real system is finite, this gives rise to a finite-size effect that can  significantly affect the signature rate of a QDS system. Compared with other protocols, the MDI protocol is particularly severely affected by the finite-size effect. Hence, in this paper, we analyze three parameter estimation models (SOB-PE, SMB1-PE, SMB2-PE) with respect to the finite-size effects on MDI-QDS. Furthermore, we conduct corresponding numerical simulations with comprehensive parameter optimizations. In contrast to the previously widely used SOB-PE model, the proposed SMB1-PE and SMB2-PE models exhibit outstanding results in terms of signature rate and transmission distance, especially the SMB1-PE model.

\section{Theoretical Models}

\subsection{Protocol procedure}
\begin{figure}[htp]
	\centering
	\includegraphics[scale=0.45]{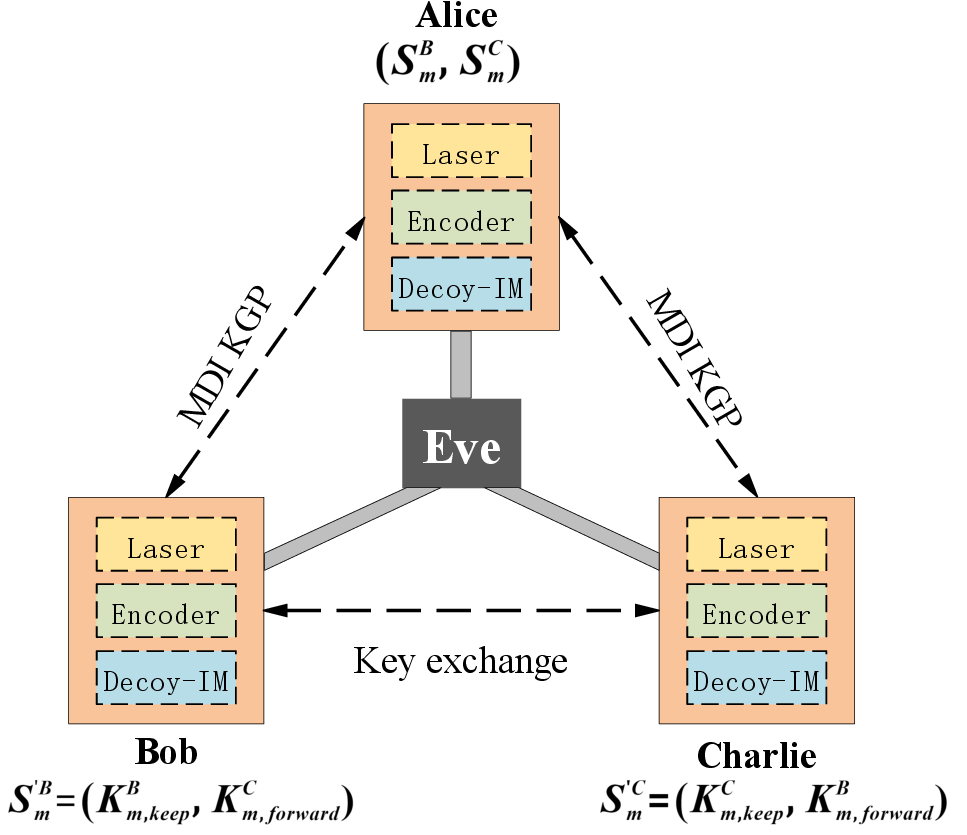}
	\caption{Schematic of the MDI-QDS protocol. Alice-Bob and Alice-Charlie perform MDI-KGP separately to generate keys. In the MDI-KGP, Alice, Bob and Charlie use the encoder and the intensity modulator (IM) to prepare quantum signals. Then the signals are sent to an untrusted party Eve, who performs a Bell state measurement. The channels between Alice-Eve, Bob-Eve, Charlie-Eve are quantum channels.}
	\label{fig1}
	\vspace{-1em}
\end{figure}

The QDS protocols are typically comprised of two stages: the first is the distribution stage, where Alice-Bob and Alice-Charlie independently perform a key-generation protocol (KGP) to generate the correlated bit strings for the signature; the second is the messaging stage, during which the message is signed and verified. The schematic diagram of MDI-QDS is shown in Fig. \ref{fig1}.

\emph{\textbf{Distribution stage:}} (1) Alice-Bob and Alice-Charlie individually generate $N$ pulses and code them using the encoder and intensity modulator (IM). Each pulse is randomly prepared with an intensity of $a\in(a_s,a_{d1},a_{d2})$ in the $X$ or $Z$ basis, and the corresponding intensity selection probabilities are $P_{a_s}, P_{a_{d1}}, P_{a_{d2}}=1-P_{a_s}-P_{a_{d1}}$, respectively. 

(2) Eve conducts an interference measurement on the received pulses and records the successful click events. Subsequently, she broadcasts the successful clicks.

(3) Alice-Bob and Alice-Charlie publicly announce their basis selection via the authenticated classical channel. The unsuccessful measurement results of mismatched basis are abandoned. Alice-Bob and Alice-Charlie use the data on $Z$ windows to extract sifted keys and the data on $X$ windows to estimate parameters. In addition, they randomly sacrifice a small number of bits on the $Z$ windows for an error rate test, leaving the remainder as a signature key pool.

(4) For each possible future message $m=0$ or $m=1$, Alice and Bob (Charlie) select bit strings from the key pool to form the signature sequence $S_m^B$ and $K_m^B$ ($S_m^C$ and $K_m^C$), where $S_m^B$ and $S_m^C$ for Alice, and $K_m^B$ ($K_m^C$) for Bob (Charlie), each of length $L$.

(5) Bob and Charlie randomly choose half of their own key bits to exchange through the certified classical channel to symmetrize their bit strings $K_m^B$ and $K_m^C$. The half of the keys that Bob (Charlie) keeps is denoted as $K_{m,keep}^B$ ($K_{m,keep}^C$), and the other half as $K_{m,forward}^B$ ($K_{m,forward}^C$). Thus, the symmetric keys held by Bob and Charlie are $S_m^{'B}=(K_{m,keep}^B,K_{m,forward}^C)$ and $S_m^{'C}=(K_{m,keep}^C, K_{m,forward}^B)$ of length $L$, respectively.

\emph{\textbf{Messaging stage:}} (6) Alice sends the message $m$ along with the signature $Sig_m=(S_m^B, S_m^C)$ to a recipient such as Bob.

(7) Bob compares his $S_m^{'B}$ with $(m,Sig_m)$ and records the number of mismatches. If the mismatch rate is fewer than $s_a$, Bob accepts the message and goes to the next step. Otherwise, he rejects the message and aborts this round. Here, $s_a$ is the authentication threshold associated with the security level of the QDS protocol.

(8) Bob forwards $(m,Sig_m)$ he received from Alice to Charlie.

(9) Charlie also checks the forwarded message in the same way. If the mismatch rate is below $s_v$ ($0<s_a< s_v<1/2$), Charlie accepts the forwarded message; otherwise Charlie rejects the message.

In MDI-QDS, the min-entropy resulting from single photon components in the half of keys kept by Bob or Charlie at the presence of Eve is
\begin{equation}
H^{\epsilon}_{min}(K^U_{m,keep}|E)\mathop{>}_{\approx}\underline{n}_{L,1}[1-H_2(\overline{e}_{L,1})],
\end{equation}
\noindent where $\epsilon$ is the security level, $U\in\left\{B,C\right\}$ denotes user Bob or Charlie, $E$ refers to the system of Eve, $\underline{n}_{L,1}$ and $\overline{e}_{L,1}$ represent the lower bound of single-photon pair counts and the upper bound of the single-photon pair error rate in $K^U_{m,keep}$, respectively. $H_2(x)=-xlog_2(x)-(1-x)log_2(1-x)$ is the binary Shannon entropy function. The minimum rate $p_E$ at which Eve can introduce errors in $K^U_{m,keep}$ can be evaluated by
\begin{equation}
H_2(p_E)=2\underline{n}_{L,1}/L[1-H_2(\overline{e}_{L,1})],
\end{equation}
where $L$ is the length of a basic block on $Z$ basis to sign message $m$. 

Through the error rate of test keys $E_{test}$, we can estimate the error rate in $K^U_{m,keep}$ with the Serfling inequality \cite{Serf} :
\begin{equation}
E_{keep}^{U}\leq E_{test}^U+\frac{2}{L}\sqrt{\frac{(\frac{L}{2}+1)(\frac{L}{2}+n_{test})ln(\frac{1}{\epsilon_{PE}})}{2n_{test}}},
\end{equation}
\noindent expect with a failure probability $\epsilon_{PE}$ and $\overline{E}_{keep}=max\left\{E_{keep}^B,E_{keep}^C\right\}$. As a result, we can calculate the threshold $s_a=E_{keep}^{U}+(p_E-E_{keep}^{U})/3$ and $s_v=E_{keep}^{U}+2(p_E-E_{keep}^{U})/3$. 

The security of digital signatures is mainly reflected in three aspects, i.e. the robustness (the probability of an honest run aborting), security against forging (the probability that a recipient generates a signature, not originating from Alice, that is accepted as authentic), and repudiation (the probability that Alice generates a signature that is accepted by Bob but then when forwarded, is rejected by Charlie) \cite{BB84QDS}. The probability of robustness, repudiation, and forging are expressed as follows:
\begin{equation}
P(Robust)\le2\epsilon_{PE},
\end{equation}
\begin{equation}
P(Repudiation)\le2exp[-\frac{1}{4}(s_v-s_a)^2L],
\end{equation}
\begin{equation}
P(Forge)\le p_F+g+\epsilon_{PE}+\varepsilon_{\underline{n}_{L,1}}+\varepsilon_{\overline{e}_{L,1}},
\end{equation}
where $g$ and $p_F$ are associated with the probability of Bob finding a signature with an error rate smaller than $s_v$, $\varepsilon_{\underline{n}_{L,1}}$ and $\varepsilon_{\overline{e}_{L,1}}$ are the error probabilities associated with the estimation of $\underline{n}_{L,1}$, and $\overline{e}_{L,1}$, respectively. Then, we can get the security level $\epsilon$ of the protocol as
\begin{equation}
max\left\{P(Robust),P(Forge),P(Repudiation)\right\}\le \epsilon.
\end{equation}


Finally, we can calculate the signature rate $(bit/pulse)$ by
\begin{equation}
R=\frac{n_{bits}}{N}.
\end{equation}
\noindent where $n_{bits}$ and $N$ are the signed bits and total number of pulse pairs, respectively.

Next, we mainly analyze and compare three parameters estimation methods for finite-size analysis of MDI-QDS.

\subsection{Parameter estimation model based on signing one bit}
\begin{figure}[htp]
	\centering
	\includegraphics[scale=0.35]{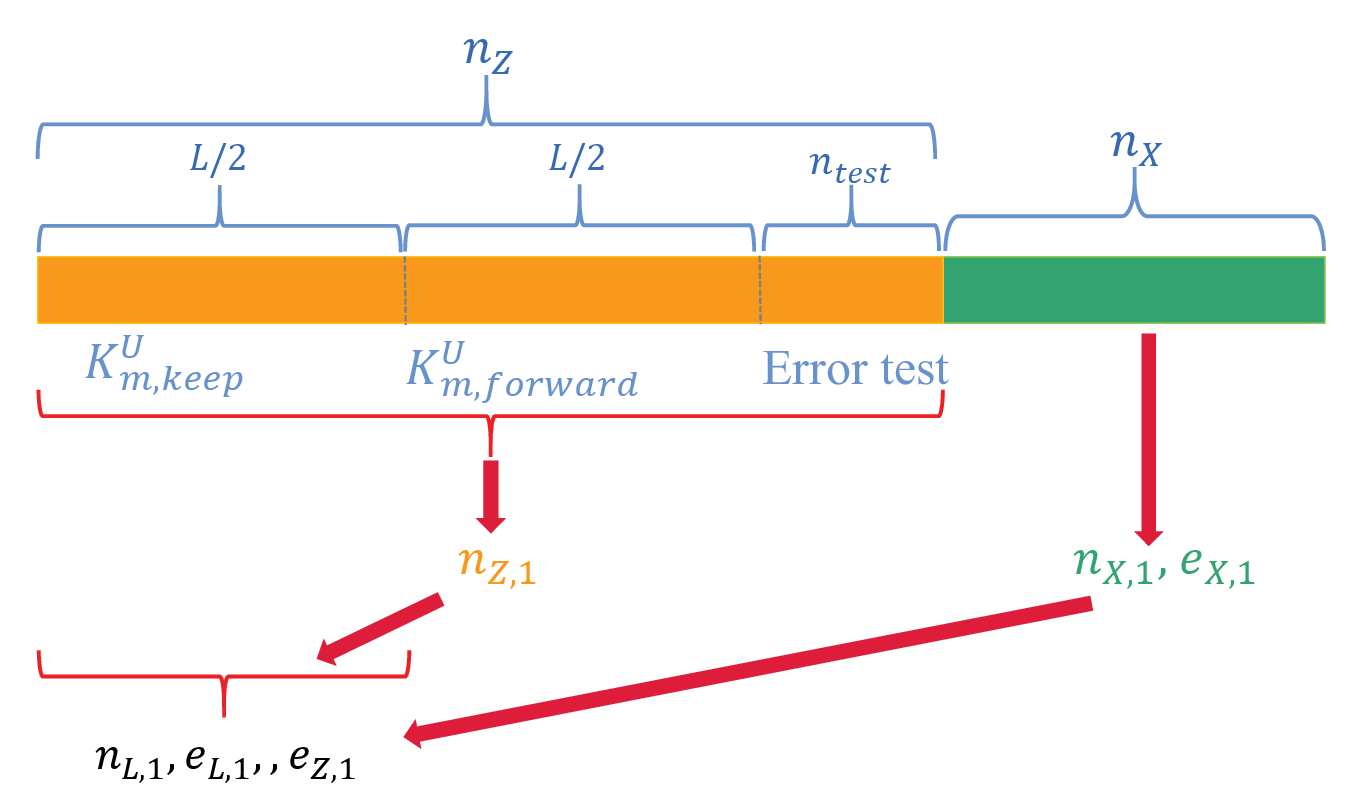}
	\caption{The relationships of different data blocks and the route of estimating parameters in the SOB-PE model. $n_X$ and $n_Z$ are the lengths of the data on $X$ basis and on $Z$ basis, $n_{test}$ is the length of the keys used for error test. $n_{X,1}$ and $e_{X,1}$ are the single-photon pair counts and single-photon pair error rate in $n_X$, while $n_{Z,1}$ and $e_{Z,1}$ are the quantities in $n_Z$, and ${n}_{L,1}$ and ${e}_{L,1}$ are the quantities in $K^U_{m,keep}$. }
	\label{fig2}
	\vspace{-1em}
\end{figure}

The parameter estimation model based on signing one bit (SOB-PE) method is commonly adopted in previous works \cite{BB84QDS,MDIQDS}, where the main idea is to sign only one bit at one time with the data of parameter estimation. In the SOB-PE model, the keys on $Z$ basis is only used to estimate single-photon counts $n_{Z,1}$ except for a part used for error testing, while the data on $X$ basis is mainly used to estimate the single-photon counts $n_{X,1} $ and error rate $e_{X,1}$. The detailed calculation regarding $n_{Z,1} $, $n_{X,1}$ and $e_{X,1}$ are provided in Appendix A. Subsequently, the corresponding single-photon quantities ($n_{L,1} $, and $e_{L,1} $) in $K^U_{m,keep}$ are estimated by the $n_{Z,1} $, $n_{X,1} $ and $e_{X,1}$. The procedure of SOB-PE is illustrated in Fig.2, which visually shows the relationships of different data blocks and relevant estimated parameters in sifted keys. Thereafter, the signed bits can be defined as

\begin{equation}
n_{bits}=\frac{N}{N_s},
\end{equation}
\noindent where $N_s$ is the number of pulse pairs used for signing one bit. As the kept keys on $Z$ basis is only used to signature one bit at a time, the signature rate $R$ $(bit/pulse)$ can be expressed as
\begin{equation}
R=\frac{1}{N_s}.
\end{equation}

\subsection{Parameter estimation model 1 based on signing multiple bits}
\begin{figure}[htbp]
	\centering
	\includegraphics[scale=0.35]{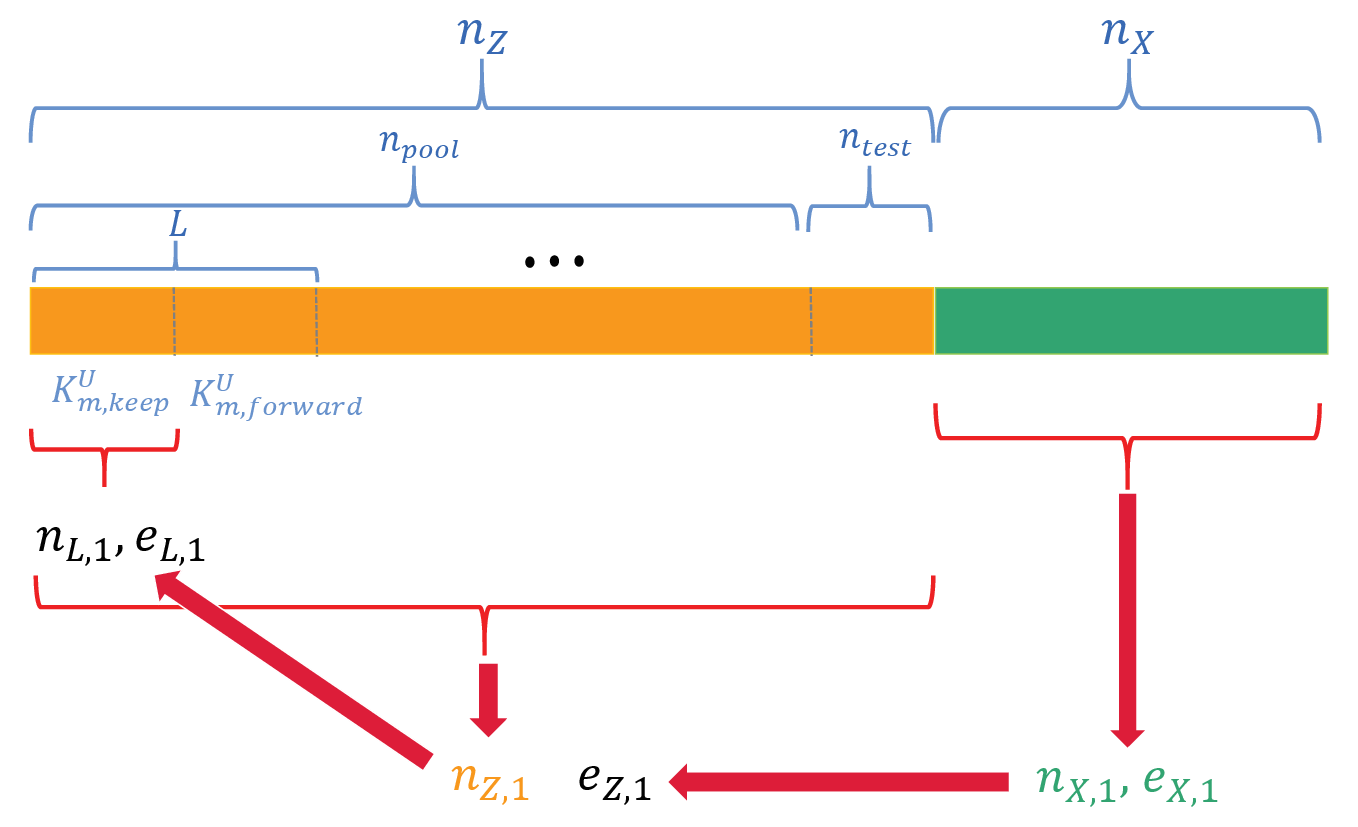}
	\caption{The relationships of different data blocks and the route of estimating parameters in the SMB1-PE model. $n_{pool}$ is the length of key pool.}
	\label{fig4}
	\vspace{-1em}
\end{figure}

Figure 3 illustrates the process of the parameter estimation model 1 based on signing multiple bits (SMB1-PE), where the kept keys on $Z$ basis can be used to sign multiple bits consecutively. The quantity $e_{Z,1}$ is defined as


\begin{equation}
e_{Z,1}=\min\left\{
\begin{array}{l}
\lceil n_{Z,1}\frac{{e}_{X,1}}{{n}_{X,1}}+(n_{Z,1}+{n}_{X,1})\times\gamma(n_{Z,1},\\
{n}_{X,1},\varepsilon^{'''}_{k,e})\rceil,n_{Z,1}
\end{array}
\right\}
\end{equation}

expect with error probability $\varepsilon_{k,e}\le\varepsilon^{'}_{k,e}+\varepsilon^{''}_{k,e}+\varepsilon^{'''}_{k,e}$, $\gamma(x,y,z)=\sqrt{(x+1)ln(z^{-1}/[2y(x+y)])}$, $\varepsilon^{'}_{k,e}$ and $\varepsilon^{''}_{k,e}$ represent their associated error probabilities, respectively. Then we can use $n_{Z,1}$ and $e_{Z,1}$ to estimate $n_{L,1}$ and $e_{L,1}$ as
\begin{equation}
n_{L,1}\ge \underline{n}_{L,1}=n_{Z,1}\frac{L}{2|Z^{a_{s},b_{s}}|}-\Lambda(|Z^{a_{s},b_{s}}|,\frac{L}{2},\epsilon_{SF}),
\end{equation}
\begin{equation}
e_{L,1}\le \overline{e}_{L,1}=e_{Z,1}+\frac{1}{\underline{n}_{L,1}}\Lambda(n_{Z,1},\underline{n}_{L,1},\epsilon_{SF}).
\end{equation}
with confidence $\epsilon_{SF}$, where $|Z^{a_{s},b_{s}}|$ is the counts when Alice and Bob select the signal state intensities $a_s$ and $b_s$ and the $Z$ basis, and $\Lambda(x,y,z)=\sqrt{(x-y+1)yln(z^{-1})/(2x)}$.

%
%

Because of the different signature modes, the way to calculate signed bits of the SMB1-PE models is different from the SOB-PE model. Subsequently, we can calculate the number of required to sign $(L)$ and the number of bits to be signed $n_{bits}$ with $n_{pool}$ keys. Finally, the $n_{bits}$ and $R$ can be described as follows:

\begin{equation}
n_{bits}=\frac{n_{pool}}{2L},
\end{equation}

\begin{equation}
R=\frac{n_{pool}}{2NL}.
\end{equation}

\subsection{Parameter estimation model 2 based on signing multiple bits}
\begin{figure}[htbp]
	\centering
	\includegraphics[scale=0.35]{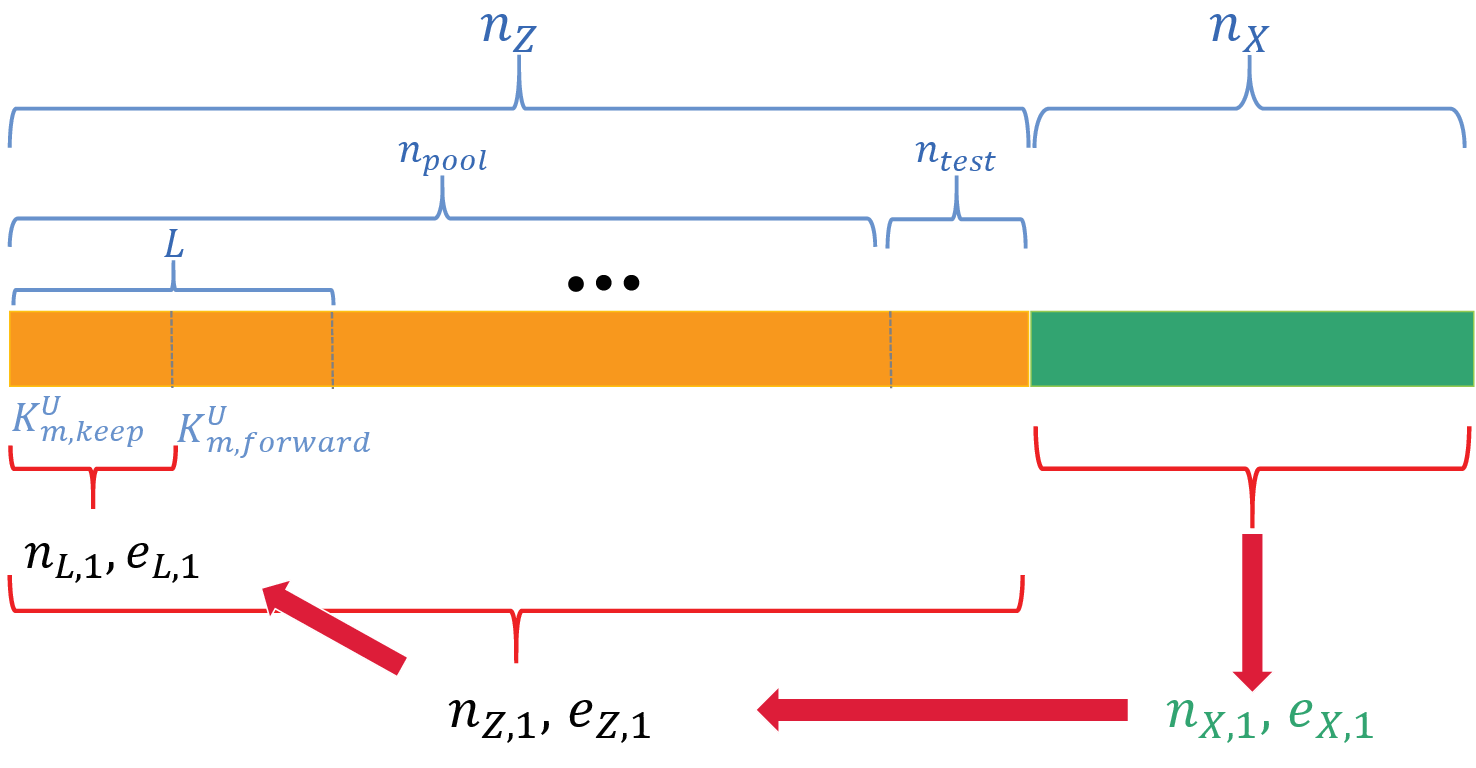}
	\caption{The relationships of different data blocks and the route of estimating parameters in the SMB2-PE model. }
	\label{fig3}
	\vspace{-1em}
\end{figure}

In the parameter estimation model 2 based on signing multiple bits (SMB2-PE), the relationships between data blocks and how to estimate parameters are shown in Fig.4. The signature mode is the same as the SMB1-PE model, the kept keys on $Z$ basis can signature multiple bits consecutively. Differently from the SMB1-PE model, we use $n_{X,1}$ to estimate the corresponding quantity $n_{Z,1}$ using the Serfling inequality \cite{Serf}. The population for single photon preparations on $Z$ basis and $X$ basis is lower bounded by

\begin{equation}
N^-_{Z,1}=(a_s+b_s)e^{-a_s-b_s}N_{z,a_sb_s}-g(N_{z,a_sb_s},\epsilon_{SF})],
\end{equation}
\begin{equation}
N^+_{X,1}=\sum_{a,b}[(a+b)e^{-a-b}N_{x,ab}+g(N_{x,ab},\epsilon_{SF})],
\end{equation}
\noindent with confidence $1-9
\epsilon_{SF}$, where $N_{z,a_sb_s}=P_{z,a_sb_s}N$ represents the pulse number of Alice choosing signal state intensity $a_s$ and Bob choosing signal state intensity $b_s$ on $Z$ basis, $N_{x,ab}=P_{x,ab}N$ means the pulse number of Alice choosing intensity $a$ and Bob choosing intensity $b$ on $X$ basis, $g(x,y)=\sqrt{2xln(y^{-1})}$ is the fluctuation in Hoeffding's inequality\cite{Hoeff}. Then we use the Serfling inequality to estimate 
\begin{equation}	
n_{Z,1}=n_{X,1}\frac{N^-_{Z,1}}{N^+_{X,1}}-\gamma(N^-_{Z,1},N^+_{X,1},\epsilon_{SF}),
\end{equation}
\noindent where $\gamma(x,y,z)=\sqrt{(x+1)(x+y)ln(z^{-1})/(2y)}$. Finally, the signature rate can also be obtained as in Eqs. (11)-(15).

In general, the commonality of the three models lies in estimating ${e_{Z,1}}$ through ${n_{Z,1}}$, ${n_{X,1}}$, and ${e_{X,1}}$. The disparities can be categorized into two aspects: first, with regard to parameter estimation, ${n_{Z,1}}$ of the SOB-PE and SMB1-PE models is directly calculated by the data on $Z$ basis, while ${n_{Z,1}}$ of the SMB2-PE model is mediately estimated by the single-photon counts on $X$ basis; second, in terms of signature mode, the SOB-PE model signs only one bit at a time, whereas the SMB1-PE and SMB2-PE models are capable of signing multiple bits.

\section{Simulation}
In this section, we perform extensive simulations to evaluate the above three parameter estimation models in two-decoy MDI-QDS. The weak decoy state intensity $a_{d2}$ is set as $5\times10^{-4}$. The remaining simulation parameters are shown in Table 1. During the course of our simulations, we carry out the full parameter optimization by employing the local search algorithm (LSA) \cite{LSA}, which encompasses the coordinate descent (CD) and backtrack search (BS) algorithms. In addition, we set the security level to $\epsilon=10^{-5}$.
\begin{table*}[htbp]
	\centering
	\centering  \caption{System parameters used in numerical simulations. $\alpha$ is the loss coefficient of fiber at telecommunication wavelength, $\eta_d$ and $P_{dc}$ are the detection efficiency and the dark count rate of detectors, respectively; $e_d$ is the optical misalignment error; $r_{ET}$ is the ratio of keys used for the error test; $\epsilon_{PE}$ and $\epsilon_{SF}$ are the failure probability of the error test and statistical fluctuation, respectively;  $g$ is probability that Bob makes $s_vL/2$ errors.}	
	\renewcommand{\arraystretch}{1.3}
	\begin{tabularx}{\linewidth}{XXXXXXXXXX}  \hline
		$\alpha$ & $\eta_d$ &  $P_{dc}$ & $e_d$ &  $r_{ET}$ &  $\epsilon_{PE}$  & $\epsilon_{SF}$ & g \\ \hline
		$0.2$ $dB/km$ & 50\% &  $10^{-7}$ & 0.03 & 5.5\% & $10^{-12}$  &  $10^{-12}$ & $10^{-12}$\\ \hline
	\end{tabularx}
	
	\label{tab1}
\end{table*}

\begin{figure}[htp]
	\centering
	\includegraphics[scale=0.35]{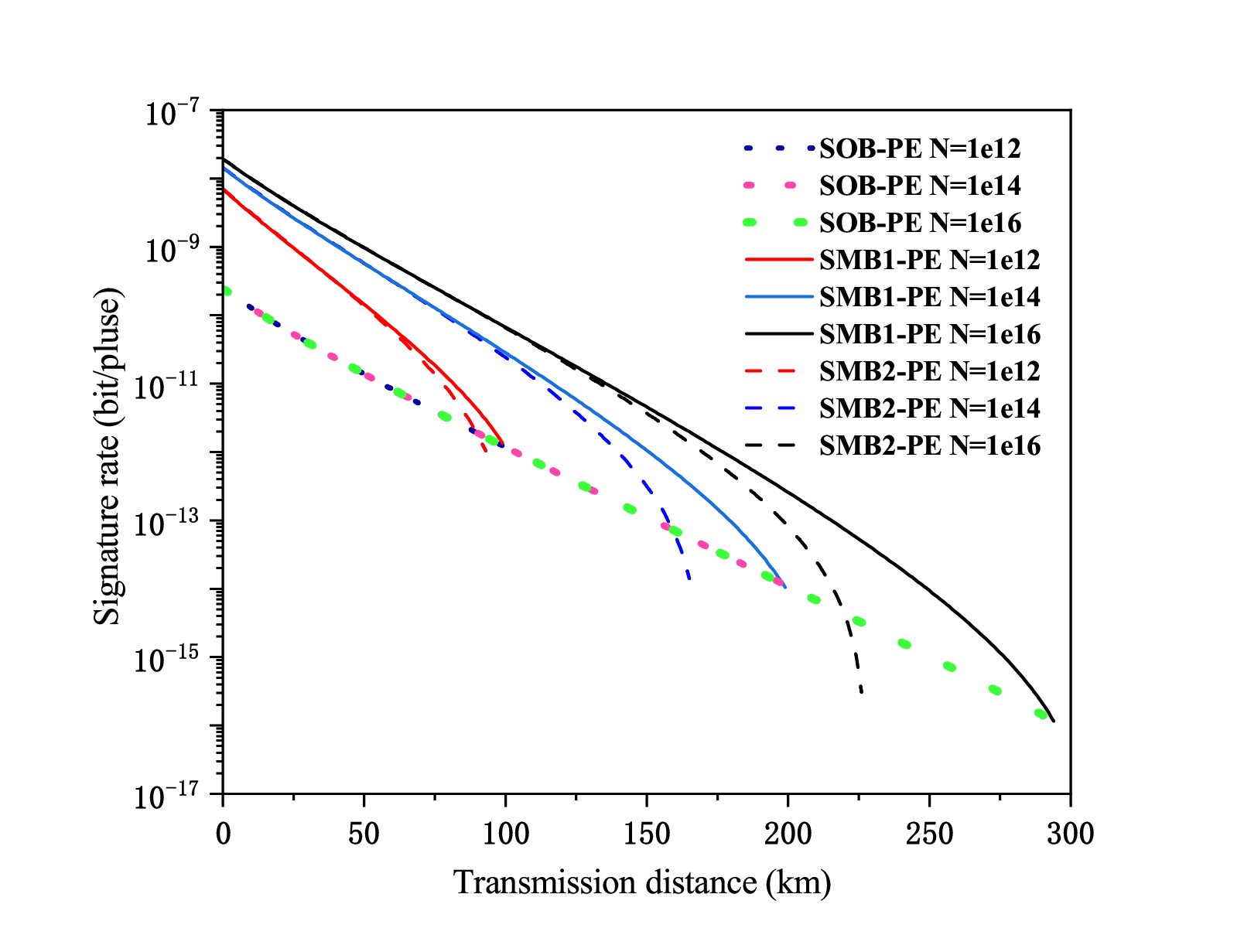}
	\caption{The signature rate of three parameter estimation models versus the transmission distance when the number of pulse pairs is $N=10^{12}, 10^{14}, 10^{16}$.}
	\label{fig5}
	
\end{figure}
\begin{figure}[htp]
	\centering
	\includegraphics[scale=0.35]{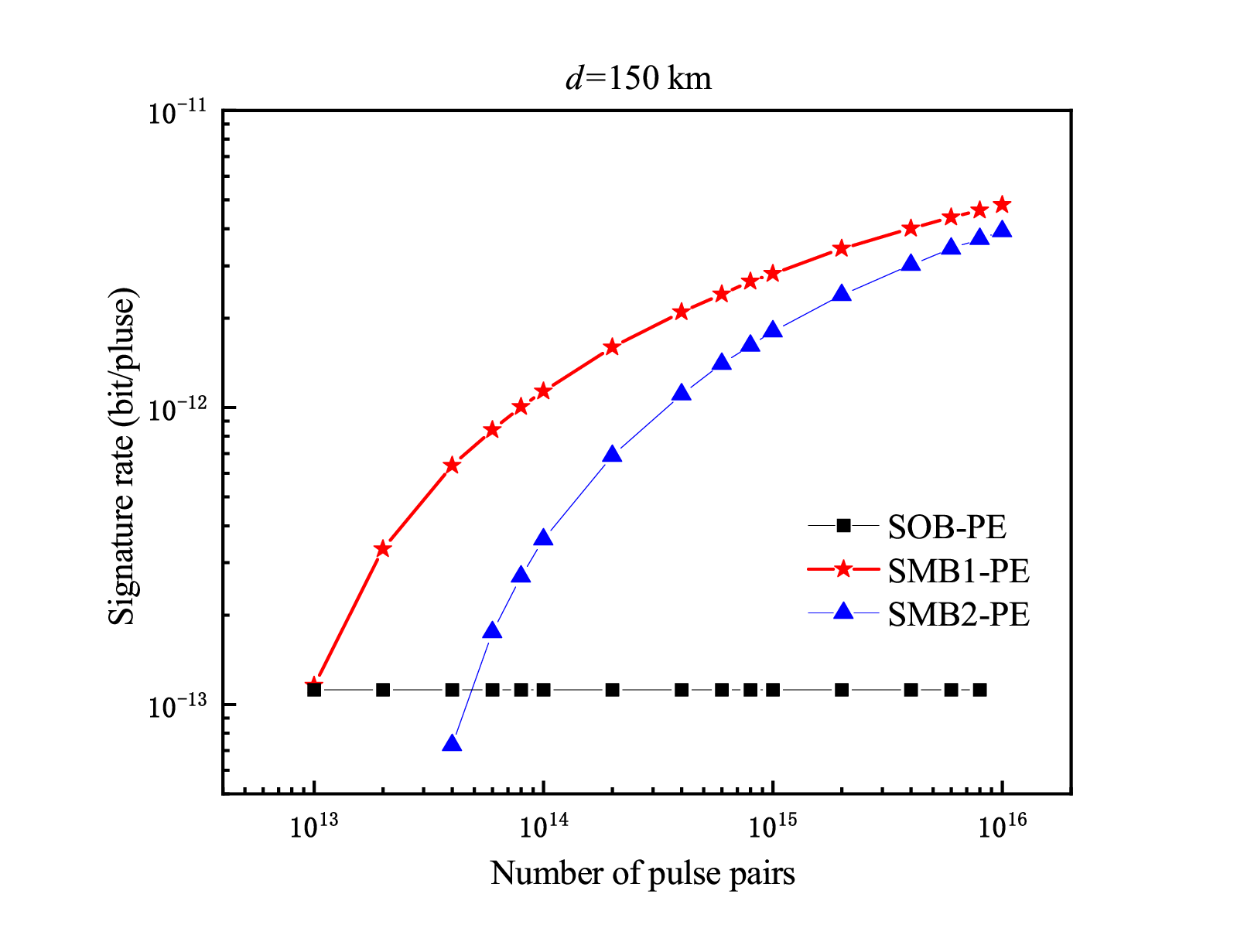}
	\caption{The signature rate of three parameter estimation models versus the number of pulse pairs at 150 km.}
	\label{fig6}
\end{figure}
The finite-size signature rates of three PE models within MDI-QDS are depicted in Fig. \ref{fig5}. Here, we use different number of pulse pairs $N=10^{12}$, $N=10^{14}$, $N=10^{16}$ to investigate the finite-size effects on the three PE models. As illustrated in Fig. 5, both of the proposed two models, namely SMB1-PE and SMB2-PE, significantly improve the performance of MDI-QDS compared to previous SOB-PE model. The SMB1-PE model maintains a higher signature rate than the SOB-PE model across all transmission distances. At the maximum transmission distance, the SMB1-PE model is equivalent to the SOB-PE model, both of which sign only one bit. The SMB2-PE model consistently outperforms the SOB-PE model in signature rate across all transmission distances except for the farthest one, at which point it lags slightly. The signature rates of the SMB1-PE and SMB2-PE models are comparable at short distances, but the SMB1-PE model's advantage becomes more evident as the distance increases. Consequently, the SMB1-PE model is the least affected by the finite-size effects and delivers the optimal performance.

In Fig. \ref{fig6}, we show the signature rate of different parameter estimation models varies with the number of pulse pairs at a distance of 150 km. The number of pulse pairs in the latter two models ranges from $10^{13}$ to $10^{16}$, while the signature rate curve of the SOB-PE model is a straight line. This is because in the first model, the kept keys on $Z$ basis is only used to sign one bit at a time, so the number of pulses used each time is the same. Upon comparison, it is evident that at a distance of 150 km, the SMB1-PE model boasts a significantly higher signature rate than the other two PE models and is least affected by the finite-size effects.

\section{Conclusion}
In conclusion, we have proposed two improved finite-size parameter estimation methods for MDI-QDS based on signing multiple bits. To evaluate the performance of three models (SOB-PE, SMB1-PE, SMB2-PE), we apply them to the two-decoy MDI-QDS. Eventually, we discover that, among all three parameter estimation models, the SMB1-PE model is particularly effective in improving the signature rate and increasing the transmission distance. This indicates that the SMB1-PE model is least affected by the finite-size effects. The proposed methods are also applicable to other QDS protocols \cite{BB84QDS,TFQDS}, and can be combined with the one-time universal hashing (OTUH) method \cite{NET2,OUTH} to further improve the signature rate. Consequently, our work provides a useful reference for the practical implementation of QDS.

\section{ACKNOWLEDGMENTS}
We also acknowledge financial support from the National Natural Science Foundation of China (12074194, 62471248), Industrial Prospect and Key Core Technology Projects of Jiangsu provincial key R\&D Program (BE2022071), and the Postgraduate Research \& Practice Innovation Program of Jiangsu Province (KYCX23\_1040).

\section{COMPETING INTERESTS}
The authors declare that there are no competing interests.

\section*{APPENDIX A: ESTIMATION OF RELEVANT PARAMETERS}
\setcounter{equation}{0}
\renewcommand{\theequation}{A\arabic{equation}}
In the Appendix we briefly discuss the estimation of $n_{Z,1}$, $n_{X,1}$, $m_{X,1}$, and $e_{X,1}$. Let $S_{W,nm}$ ($W\in\left\{X,Z\right\}$) denote the number of signals sent by Alice and Bob with $n$ and $m$ photons, respectively. Then if 
\begin{equation}
(2\varepsilon_{a,b}^{-1})^{1/\mu_{L}^{a,b}}\le exp[3/(4\sqrt2)^2],
\end{equation}
and
\begin{equation}
(\hat{\varepsilon}_{a,b}^{-1})^{1/\mu_{L}^{a,b}}\le exp(1/3),
\end{equation}
with the parameter $\mu_{L}^{a,b}$ given by
\begin{equation}
\mu_{L}^{a,b}=|W^{a,b}|-\sqrt{\sum_{a,b}|W^{a,b}|/2ln(1/\epsilon_{a,b})},
\end{equation}
this implies that
\begin{equation}
|W^{a,b}|=\sum_{n,m}p_{a,b|nm,Z}S_{W,nm}+\delta_{a,b},
\end{equation}
\noindent
expect with error probability $\varepsilon^{'}_{a,b}=\epsilon_{a,b}+\varepsilon_{a,b}+\hat{\varepsilon}_{a,b}$. Here, $p_{a,b|nm,Z}$ means the conditional probability that Alice and Bob have selected the intensity $a$ and $b$, respectively, given that their signals contain $n$ and $m$ photons, respectively, prepared on $Z$ basis. And $\delta_{a,b}=g(\sum_{n,m}p_{a,b|nm,Z}S_{W,nm},\varepsilon^{'}_{a,b})$, $g(x,y)=\sqrt{2xln(y^{-1})}$. Then the quantity $n_{Z,1}$ and $n_{X,1}$ can be written as
\begin{equation}
n_{Z,1}=p_{a_s,b_s|11,Z}S_{Z,11}-\delta_{Z,1},
\end{equation}

\begin{equation}
n_{X,1}=\sum_{a,b}p_{a,b|11,X}S_{X,11}-\delta_{X,1},
\end{equation}

\noindent  expect with error probability $\varepsilon_1$ where $\delta_{Z,1}=g(p_{a_s,b_s|11,Z}S_{Z,11},\varepsilon_1)$,  $\delta_{X,1}=g(\sum_{a,b}p_{a,b|11,X}S_{X,11},\varepsilon_1)$, and $p_{a,b|nm,W}$ means the conditional probability that Alice and Bob have selected the intensity $a$ and $b$, given that their signals contain $n$ and $m$ photons, prepared on the $W$ basis.

A similar approach is followed to estimate $m_{X,1}$ as


\begin{equation}
m_{X,1}=\sum_{a,b}p_{a,b|11,X}E_{11}-\delta_1,
\end{equation}
\noindent expect with error probability $\varepsilon_1$ where $\delta_1=g(\sum_{a,b}p_{a,b|11,X}E_{11},\varepsilon_1)$, and $E_{nm}$ denote the number of signals sent by Alice and Bob with single-photon pair on the $X$ basis, and the bits of Ailce and Bob differ after applying the bit flip operation. Then the quantity $e_{X,1}$ can be expressed as

\begin{equation}
e_{X,1}=\frac{m_{X,1}}{n_{X,1}}.
\end{equation}


\begin{thebibliography}{1}
	\bibliographystyle{IEEEtran}
	\bibitem{APP1}
	X.-Y. Cao, B.-H. Li, Y. Wang, Y. Fu, H.-L. Yin, and Z.-B. Chen, "Experimental quantum e-commerce", \textit{Sci. Adv.} vol. 10, eadk3258, 2024.
	
	\bibitem{APP2}
	P. Schiansky, J. Kalb, E. Sztatecsny, M.-C. Roehsner, T. Guggemos, A. Trenti, M. Bozzio, and P. Walther, "Demonstration of quantum-digital payments", \textit{Nat. Commun.}, vol. 14, pp. 3849, 2023.
	
	
	\bibitem{Eve}
	N. Wolchover, “A tricky path to quantum-safe encryption,” \textit{Quanta Mag.,} Sept. 2015.
	\bibitem{Eve1}
	R. Amiri and E. Andersson, "Unconditionally secure quantum signatures", \textit{Entropy}, vol. 17, pp. 5635, 2015.
	
	
	\bibitem{QKD1}
	W.-K. Wootters, W.-H. Zurek, "A single quantum cannot be cloned," \textit{Nature}, vol. 299, no. 5886, pp. 802–803, 1982.
	
	\bibitem{QKD2}
	H.-K. Lo and H.-F. Chau, “Unconditional security of quantum key distribution over arbitrarily long distances,” \textit{Science}, vol. 283, pp. 2050–2056, 1999.
	
	\bibitem{QKD3}
	V. Scarani, H. Bechmann-Pasquinucci, N.-J. Cerf, M. Dusek, N. Lutkenhaus, and M. Peev, "The security of practical quantum key distribution," \textit{Rev. Mod. Phys.}, vol. 81, no. 3, pp. 1301–1350, Sept. 2009.
	
	\bibitem{QDS2001}
	D. Gottesman, and I. Chuang, "Quantum digital signatures," \textit{ArXiv:quant-ph/0105032}, 2001.
	
	\bibitem{SEC1}
	R.-J. Collins, R.-J. Donaldson, V. Dunjko, P. Wallden, P.-J. Clarke, E. Andersson, J. Jeffers, and G.-S. Buller, "Realization of quantum digital signatures without the requirement of quantum memory", \textit{Phys. Rev. Lett.}, vol. 113, no. 5, pp. 040502, 2014.
	
	\bibitem{SEC2}
	H.-L. Yin, Y. Fu, and Z.-B. Chen, "Practical quantum digital signature", \textit{Phys. Rev. A}, vol. 93, no. 12, pp. 032316, Mar. 2016.
	
	\bibitem{SEC3}
	Y.-S. Lu, X.-Y. Cao, C.-X. Weng, J. Gu, Y.-M. Xie, M.-G. Zhou, H.-L. Yin, and Z.-B. Chen, "Efficient quantum digital signatures without symmetrization step," \textit{Opt. Express}, vol. 29, pp. 10162-10171, 2021.
	
	\bibitem{SEC4}
	C.-X. Weng, Y.-S. Lu, R.-Q. Gao, Y.-M. Xie, J. Gu, C.-L. Li, B.-H. Li, H.-L. Yin, and Z.-B. Chen, "Secure and practical multiparty quantum digital signatures," \textit{Opt. Express}, vol. 29, pp. 27661-27673, 2021.
	
	\bibitem{BB84QDS}
	R. Amiri, P. Wallden, A. Kent, and E. Andersson, "Secure quantum signatures using insecure quantum channels," \textit{Phys. Rev. A}, vol. 93, no. 9, pp. 032325, 2016.
	
	
	\bibitem{MDIQDS}
	I.-V. Puthoor, R. Amiri, P. Wallden, M. Curty, and E. Andersson, "Measurement-device-independent quantum digital signatures," \textit{Phys. Rev. A,} vol. 94, no. 11, pp. 022328, Aug. 2016
	
	\bibitem{MDIQDS2}
	J.-W. Bian, B.-H. Li, Y.-M. Xie, H.-L. Yin, and Z.-B. Chen, "Asynchronous measurement-device-independent quantum digital signatures", \textit{Phys. Rev. A,} vol. 110, no. 15, pp. 012609, Jul. 2024.
	
	\bibitem{TFQDS}
	C.-H. Zhang, X.-Y. Zhou, C.-M. Zhang, J. Li, and Q. Wang, "Twin-field quantum digital signatures," \textit{Opt. Lett.}, vol. 46, no. 15, pp. 3757-3760, Aug. 2021.
	\bibitem{TFQDS2}
	M.-H. Zhang, J.-H. Xie, J.-Y. Wu, L.-Y. Yue, C. He, Z.-W. Cao, J.-Y. Peng, "Practical long-distace twin-field quantum digital signatures," \textit{Quantum Inf. Process}, vol. 21, pp. 150, Apr. 2022.
	\bibitem{PP1}
	H. Xu, Z.-W. Yu, C. Jiang, X.-L. Hu, and X.-B. Wang, "Sending-or-not-sending twin-field quantum key distribution: Breaking the direct transmission key rate," \textit{Phys. Rev. A}, vol. 101, no. 10, pp. 042330, Apr. 2020. 
	
	\bibitem{PP2}
	J.-Q. Qin, C. Jiang, Y.-L. Yu, and X.-B. Wang, "Quantum Digital Signatures with Random Pairing," \textit{Phys. Rev. Appl.}, vol. 17, no. 15, pp. 044047, Apr. 2022. 
	
	
	\bibitem{AD1}
	U.-M. Maurer, "Secret key agreement by public discussion from common information," \textit{IEEE Trans. Inform. Theory}, vol. 39, no. 3, pp. 733–742, May 1993.
	
	
	
	\bibitem{AD5}
	K. Zhang, J. Liu, H. Ding, X. Zhou, C. Zhang, and Q. Wang, "Asymmetric measurement-device-independent quantum key distribution through advantage distillation," \textit{Entropy}, vol. 25, no. 8, pp. 1174, Aug. 2023.
	
	
	\bibitem{AD6}
	Y. Zhou, R.-Q. Wang, C.-M. Zhang, Z.-Q. Yin, Z.-H. Wang, S. Wang, W. Chen, G.-C. Guo, and Z.-F. Han, "Sending-or-not-sending twin-field quantum key distribution with advantage distillation", \textit{Phys. Rev. Appl.}, vol. 21, no. 9, pp. 014036, Jan. 2024.
	
	
	
	
	
	\bibitem{LD1}
	H.-L. Yin, Y. Fu, H. Liu, Q.-J. Tang, J. Wang, L.-X. You, W.-J. Zhang, S.-J. Chen, Z. Wang, Q. Zhang, T.-Y. Chen, Z.-B. Chen, and J.-W. Pan, "Experimental quantum digital signature over 102 km," \textit{Phys. Rev. A}, vol. 95, no. 9, pp. 032334, Mar. 2017.
	
	\bibitem{LD2}
	C.-H. Zhang, X.-Y. Zhou, H.-J. Ding, C.-M. Zhang, G.-C. Guo, and Q. Wang, "Proof-of-principle demonstration of passive decoy-state quantum digital signatures over 200 km", \textit{Phys. Rev. Appl.}, vol. 10, no. 8, pp. 034033, 2018.
	
	\bibitem{HD1}
	X.-B. An, H. Zhang, C.-M. Zhang, W. Chen, S. Wang, Z.-Q. Yin, Q. Wang, D.-Y. He, P.-L. Hao, S.-F. Liu, X.-Y. Zhou, G.-C. Guo, and Z.-F. Han, "Practical quantum digital signature with a gigahertz BB84 quantum key distribution system," \textit{Opt. Lett.}, vol. 44, pp. 139-142, 2019.
	
	\bibitem{NET3}
	G.-L. Roberts, M. Lucamarini, Z.-L. Yuan, J.-F. Dynes, L.-C. Comandar, A.-W. Sharpe, A.-J. Shields, M. Curty, I.-V. Puthoor, and E. Andersson, "Experimental measurement-device-independent quantum digital signatures", \textit{Nat. Commun.}, vol. 8, pp. 1, Oct. 2017.
	
	\bibitem{LD3}
	H.-J. Ding, J.-J. Chen, L. Ji, X.-Y. Zhou, C.-H. Zhang, C.-M. Zhang, and Q. Wang, "280-km experimental demonstration of a quantum digital signature with one decoy state," \textit{Opt. Lett.}, vol. 45, pp. 1711-1714, 2020.
	
	\bibitem{NET2}
	H.-L. Yin, Y. Fu, C.-L. Li, C.-X. Weng, B.-H. Li, J. Gu, Y.-S. Lu, S. Huang, and Z.-B. Chen, "Experimental quantum secure network with digital signatures and encryption", \textit{National Science Review}, vol. 10, Apr. 2023, nwac228.
	
	
	
	\bibitem{Chip}
	Y.-Q. Du, B.-H. Li, X. Hua, X.-Y. Cao, Z.-G. Zhao, F. Xie, Z.-R. Zhang, H.-L. Yin, X. Xiao, and K.-J. Wei, "High-rate quantum digital signatures network with integrated silicon photonics", \textit{ArXiv:quant-ph/2407.07513}, 2024.
	\bibitem{MU1}
	W. Qu, Y. Zhang, H. Liu, T. Dou, J. Wang, Z. Li, S. Yang, and H. Ma, "Multi-party ring quantum digital signatures", \textit{J. Opt. Soc. Am. B}, vol. 36, pp. 1335-1341, 2019.
	
	\bibitem{MU2}
	C.-X. Weng, Y.-S. Lu, R.-Q. Gao, Y.-M. Xie, J. Gu, C.-L. Li, B.-H. Li, H.-L. Yin, and Z.-B. Chen, "Secure and practical multiparty quantum digital signatures," \textit{Opt. Express}, vol. 29, pp. 27661-27673, 2021.
	\bibitem{MU3}
	Y. Pelet, I. V. Puthoor, N. Venkatachalam, S. Wengerowsky, M. Loncari ˇ c, S. P. Neumann, B. Liu, Željko Samec, M. Stip ´ cevi ˇ c,´R. Ursin, E. Andersson, J. G. Rarity, D. Aktas, and S. K. Joshi, "Unconditionally secure digital signatures implemented in an eight-user quantum network", \textit{New J. Phys.}, vol. 24, pp. 093038, 2022.
	
	\bibitem{SPS}
	L. Zhan, C. H. Zhang, N. Lu, X. R. Qian, H. J. Ding, J. Y. Liu, X. Y. Zhou, and Q. Wang, "Experimental quantum digital signature based on heralded single-photon sources," \textit{Quantum Inf. Process}, vol. 23, pp. 25, 2024.
	
	
	
	\bibitem{FT1}
	C. Croal, C. Peuntinger, B. Heim, I. Khan, C. Marquardt, G. Leuchs, P. Wallden, E. Andersson, and N. Korolkova, "Free-Space Quantum Signatures Using Heterodyne Measurements", \textit{Phys. Rev. Lett.}, vol. 117, no. 5, pp. 100503, Sep. 2016.
	
	\bibitem{NET1}
	H.-L. Yin, W.-L. Wang, Y.-L. Tang, Q. Zhao, H. Liu, X.-X. Sun, W.-J. Zhang, H. Li, I.-V. Puthoor, L.-X. You, E. Andersson, Z. Wang, Y. Liu, X. Jiang, X.-F. Ma, Q. Zhang, M. Curty, T.-Y. Chen, and J.-W. Pan, "Experimental measurement-device-independent quantum digital signatures over a metropolitan network", \textit{Phys. Rev. A}, vol. 95, no. 10, pp. 042338, Apr. 2017.
	
	
	
	
	
	%
	%
	%
	%
	%
	%
	%
	%
	%
	%
	%
	
	\bibitem{MDI}
	M. Curty, F.-H. Xu, W. Cui, C.-C.-W. Lim, K. Tamaki, H.-K. Lo, "Finite-key analysis for measurement-device-independent quantum key distribution," \textit{Nat. Commun.}, vol. 5, pp. 3732, Apr. 2014.
	
	\bibitem{Serf}
	R.-J. Serfling, "Probability Inequalities for the Sum in Sampling without Replacement", \textit{The Annals of Statistics,} vol. 2, no. 1, pp. 39,	1974.
	
	
	\bibitem{Hoeff}
	W. Hoeffding, "Probability inequalities for sums of bounded random variables", \textit{Journal of the American Statistical Association}, vol. 58, no. 301, pp. 13, 1963.
	
	
	\bibitem{LSA}
	F.-H. Xu, H. Xu, and H.-K. Lo, "Protocol choice and parameter optimization in decoy-state measurement-device-independent quantum key distribution", \textit{Phys. Rev. A}, vol. 89, pp. 052333, 2014.
	
	
	\bibitem{OUTH}
	B.-H. Li, Y.-M. Xie, X.-Y. Cao, C.-L. Li, Y. Fu, H.-L. Yin, and Z.-B. Chen, "One-time universal hashing quantum digital signatures without perfect keys", \textit{Phys. Rev. Appl.}, vol. 20, pp. 044011, 2023.
	
	
\end{thebibliography}
\end{document}